\documentstyle[prb,aps,graphicx]{revtex}
\RequirePackage{multicol}

\newif\ifmynarrow \mynarrowfalse

\renewcommand{\narrowtext}{%
  \ifmynarrow\hspace*{\fill}\raisebox{-1ex}[0pt][0pt]{%
    \rule{0.3pt}{1ex}%
    \rule[1ex]{20.5pc}{0.3pt}}\fi
  \mynarrowtrue
  \vspace{-1.0ex}%
  \begin{multicols}{2}%
  \par\global\columnwidth20.5pc
  \global\hsize\columnwidth\global\linewidth\columnwidth
  \global\displaywidth\columnwidth}

\renewcommand{\widetext}{%
  \end{multicols}%
  \vspace{-2.5ex}%
  \noindent\raisebox{1ex}[0pt][0pt]{%
    \rule{20.5pc}{0.3pt}%
    \rule{0.3pt}{1ex}}%
  \par\global\columnwidth42.5pc
  \global\hsize\columnwidth\global\linewidth\columnwidth
  \global\displaywidth\columnwidth}

\begin{document}

\title{Anomalous Rashba spin splitting in two-dimensional hole systems}

\draft

\author{R.~Winkler}
\address{Institut f\"ur Technische Physik III, Universit\"at
Erlangen-N\"urnberg, Staudtstr. 7, D-91058 Erlangen, Germany}

\author{H.~Noh, E.~Tutuc, and M.~Shayegan}
\address{Department of Electrical Engineering, Princeton University,
Princeton, New Jersey 08544}

\date{June 13, 2001}
\maketitle
\begin{abstract}
  It has long been assumed that the inversion asymmetry-induced
  Rashba spin splitting in two-dimensional (2D) systems at zero
  magnetic field is proportional to the electric field that
  characterizes the inversion asymmetry of the confining potential.
  Here we demonstrate, both theoretically and experimentally, that
  2D heavy hole systems in accumulation layer-like single
  heterostructures show the opposite behavior, i.e., a decreasing,
  but nonzero electric field results in an increasing Rashba
  coefficient.
\end{abstract}

\narrowtext

Spin degeneracy of electron and hole states in a solid stems from
the inversion symmetry in space and time. If the spatial inversion
symmetry is broken, then there is a splitting of the single particle
states even at magnetic field $B=0$ (Ref.\ \onlinecite{kit63}). In
quasi two-dimensional (2D) semiconductor structures, the bulk
inversion asymmetry (BIA) of the underlying crystal structure (e.g.,
a zinc blende structure), and the structure inversion asymmetry
(SIA) of the confining potential contribute to the $B=0$ spin
splitting. \cite{ros89} While BIA is fixed, the so-called Rashba
spin splitting \cite{byc84} due to SIA can be tuned by means of
external gates that change the electric field $\bf E$ in the sample.
\cite{nit97,pap99} The $B=0$ spin splitting is of significant
current interest both because of its fundamental importance as well
as its possible device applications. \cite{pasps}

For many years it has been assumed that the Rashba spin splitting in
2D systems is proportional to the electric field that characterizes
the inversion asymmetry of the confining potential.
\cite{ros89,byc84} In single heterostructures, where SIA is the
dominant source of spin splitting, the electric field is determined
by the density-dependent self-consistent potential. \cite{ste72} One
would thus expect that the spin splitting decreases with density,
although for 2D electron systems this effect may partly be
compensated by many-particle effects that tend to increase the
Rashba spin splitting for low densities. \cite{che99} Here we
demonstrate, both theoretically and experimentally, that 2D heavy
hole systems in accumulation layer-like single heterostructures show
the opposite behavior, namely, a decreasing, but nonzero electric
field results in an increasing Rashba coefficient. Contrary to
electrons, however, exchange-correlation effects in the low-density
regime decrease the spin splitting. We show that this surprising
result is a consequence of heavy hole--light hole (HH-LH) coupling
in 2D hole systems. We obtain good qualitative agreement between
calculated and measured spin splittings in 2D hole systems in GaAs
heterostructures where the density and the spin splitting are varied
by means of an external gate.  Our results are applicable to many
systems, as most III-V semiconductors have essentially the same band
structure that is underlying our investigation.

To lowest order of the wave vector $\bf k$ and electric field $\bf
E$, the SIA spin splitting of electron states in the $\Gamma_6^c$
conduction band is given by the Rashba term \cite{byc84}
\begin{equation}
\label{eq:rashbafull}
H_{6c}^{\rm SO} = \alpha \, {\bf k} \times {\bf E} \cdot
\bbox{\sigma} .
\end{equation}
Here $\bbox{\sigma} = (\sigma_x, \sigma_y, \sigma_z)$ denotes the
Pauli spin matrices and $\alpha$ is a material-specific prefactor.
\cite{ros89} We assume ${\bf E} = (0,0,E_z)$. Treating the
off-diagonal $\bf k\cdot p$ coupling between electron and hole
states by third order L\"owdin perturbation theory, \cite{bir74} we
obtain for the Rashba coefficient $\alpha_\lambda$ of the lowest
electron subband $\lambda=1$
\begin{equation}
\label{eq:rash_6c_sub_par}
\alpha_1 = e\,P^2 \, a
\bigg( \frac{1}{\Delta_{11}^{cl}} \frac{1}{\Delta_{12}^{cl}} 
     - \frac{1}{\Delta_{11}^{cs}} \frac{1}{\Delta_{12}^{cs}} \bigg),
\end{equation}
where $P$ is Kane's momentum matrix element \cite{tre79} and
$\Delta_{\lambda\lambda'}^{\nu\nu'} \equiv {\cal E}_\lambda^\nu -
{\cal E}_{\lambda'}^{\nu'}$ with ${\cal E}_\lambda^c$, ${\cal
E}_\lambda^h$, ${\cal E}_\lambda^l$, and ${\cal E}_\lambda^s$ the
energy of the $\lambda$th electron, HH, LH, and split-off subband,
respectively. \cite{iv_rash} The numerical prefactor $a$ depends on
the geometry of the confining quantum well (QW). In an infinitely
deep rectangular QW we have $a=256/(81\pi^2)$. According to Eq.\ 
(\ref{eq:rashbafull}) we obtain a spin splitting $\pm \alpha E_z \,
k_\|$ of the subband dispersion ${\cal E} ({\bf k}_\|)$ that is
proportional to the electric field $E_z$ and is linear in the
in-plane wave vector ${\bf k}_\| = (k_x,k_y,0)$. A detailed analysis
reveals that spin splitting of electron states depends on the
electric field $E_v$ in the valence band that differs from the
electric field $E_c$ in the conduction band by the contributions of
the interfaces. \cite{las85,and97} However, the important point here
is that in a single heterostructure both $E_v$ and $E_c$ are
determined by the self-consistent Hartree potential.

For hole systems in the $\Gamma_8^v$ valence band (point group
$T_d$), the dominant contribution to Rashba spin splitting is given
by the term \cite{win00a}
\begin{equation}
\label{eq:rash_8v}
H_{8v}^{\rm SO} = \beta \, {\bf k} \times {\bf E} \cdot {\bf J},
\end{equation}
where $\beta$ is a system-dependent prefactor and ${\bf J} =
(J_x,J_y,J_z)$ are the angular momentum matrices for $j=3/2$. We
neglect here the small corrections in $H_{8v}^{\rm SO}$ due to the
$\bf k\cdot p$ coupling to remote bands such as the higher
$\Gamma_8^c$ and $\Gamma_7^c$ conduction bands. \cite{koster}
Quantum confinement reduces the symmetry from the cubic point group
$T_d$ to $D_{2d}$. Therefore, the four-fold degeneracy of the
$\Gamma_8^v$ band ($T_d$) is lifted and we obtain two-fold
degenerate subspaces transforming according to $\Gamma_7$ and
$\Gamma_6$ of the point group $D_{2d}$ (Refs.\ 
\onlinecite{jor90,pgroup}). The irreducible representation
$\Gamma_7$ corresponds to the LH states ($z$~component of angular
momentum $j_z = \pm 1/2$) whereas $\Gamma_6$ corresponds to the HH
states ($j_z = \pm 3/2$). To lowest order in $k$ the effective
Rashba Hamiltonian for the LH states is the same as in Eq.\ 
(\ref{eq:rashbafull}) for electron states. \cite{ref_ele} For HH
states, however, spin splitting is mediated by a coupling to the LH
states so that, to lowest order, spin splitting of HH states is of
third order in $\bf k$ (Ref.~\onlinecite{win00a}). Neglecting
anisotropic corrections in the Hamiltonian we have
\begin{equation}
\label{eq:rash_HH}
H_h^{\rm SO} = \beta^h \, E_z 
(\sigma_+ k_-^3 + \sigma_- k_+^3),
\end{equation}
with $\sigma_\pm = 1/2(\sigma_x \pm i \sigma_y)$ and $k_\pm = k_x
\pm i k_y$. Treating the off-diagonal HH-LH coupling by third order
L\"owdin perturbation theory \cite{bir74} we obtain for the Rashba
coefficient $\beta^h_\lambda$ of the lowest HH subband $\lambda=1$
\begin{equation}
\label{eq:rash_HH_fak}
\beta^h_1 = ia\, \gamma_3 (\gamma_2+\gamma_3)
\bigg[ \frac{1}{\Delta_{11}^{hl}} 
 \bigg(\frac{1}{\Delta_{12}^{hl}} 
     - \frac{1}{\Delta_{12}^{hh}} \bigg)
     + \frac{1}{\Delta_{12}^{hl} \; \Delta_{12}^{hh}} \bigg]
\end{equation}
where $\gamma_2$ and $\gamma_3$ are the Luttinger parameters.
\cite{lut56} In an infinitely deep rectangular QW we have
$a=64/(9\pi^2)$. We see from Eq.\ (\ref{eq:rash_HH_fak}) that the
Rashba spin splitting of HH states depends not only on the electric
field $E_z$ but also on the separation between the HH and LH
subbands. A decreasing separation gives rise to an increasing Rashba
coefficient $\beta^h_1$. The factor $\gamma_3 (\gamma_2+\gamma_3)$
in Eq.\ (\ref{eq:rash_HH_fak}) refers to a quantum structure grown
in the crystallographic direction [001]. The expressions for other
growth directions are similar, but the other terms in Eq.\ 
(\ref{eq:rash_HH_fak}) remain unchanged. We remark that for typical
hole densities only the lowest HH subband is occupied.

The electric field $E_z$ that enters into the Rashba Hamiltonian
depends on the charges in the system. We will show now that
accumulation layer-like single heterostructures behave rather
differently with respect to changes of the 2D charge density as
compared to other quasi 2D semiconductor structures.

In a rectangular QW, a small density $N$ and a small asymmetry imply
that the properties of the system are controlled by the effective
potential steps at the interfaces, i.e., changes in $N$ or $E_z$
have a minor effect in this regime. In an inversion layer-like
heterostructure, we always have a band bending of the order of the
fundamental gap so that, for small densities, the Hartree potential
and $E_z$ are determined by the space charges due to the given
concentration of ionized majority impurities in the system. For
accumulation layer-like systems, on the other hand, it was shown by
Stern, \cite{ste74} that the space charge layer is controlled by the
much smaller concentration of minority impurities in the system.
Thus, even for a small 2D density, the dominant contribution to the
Hartree potential stems from the charges in the 2D system itself.
Therefore, over a wide range of densities $N$, the electric field
$E_z$ is proportional to $N$. In single heterostructures, the
subband separations are approximately proportional to $E_z$. Using
the triangular well approximation \cite{ste72} we have, for the
subband energies ${\cal E}_\lambda^\nu$ measured from the
corresponding bulk band edge, ${\cal E}_\lambda^\nu \propto
E_z^{2/3}$ which implies ${\cal E}_\lambda^\nu \propto N^{2/3}$ and
$\beta^h_\lambda \propto N^{-4/3}$. Therefore, we can expect from
Eqs.\ (\ref{eq:rash_HH}) and (\ref{eq:rash_HH_fak}) that
accumulation layer-like 2D HH systems show a Rashba spin splitting
that increases when $N$ and $E_z$ are reduced. On the other hand,
the coefficient (\ref{eq:rash_6c_sub_par}) is essentially
independent of $N$ and $E_z$ because the energy gaps
$\Delta_{\lambda\lambda'}^{\nu\nu'}$ are always of the order of the
fundamental gap. \cite{iv_rash}

In order to validate these qualitative arguments we present next the
results of realistic, fully self-consistent subband calculations.
\cite{win93a} We use an $8\times 8$ multiband Hamiltonian
\cite{tre79} that includes the lowest conduction band $\Gamma_6^c$,
the topmost valence band $\Gamma_8^v$ and the split-off valence band
$\Gamma_7^v$. The simpler $4\times 4$ Luttinger Hamiltonian,
\cite{lut56} taking into account only the band $\Gamma_8^v$, gives
essentially the same results. We have checked that higher conduction
bands have a minor influence. Many-particle effects are taken into
account based on a density-functional approach. \cite{ste84,dft}
From these calculations we obtain the difference $\Delta N = N_+ -
N_-$ between the spin subband densities $N_\pm$ as a function of the
total density $N = N_+ + N_-$.

\begin{figure}
\centerline{\includegraphics[width=0.67\columnwidth]{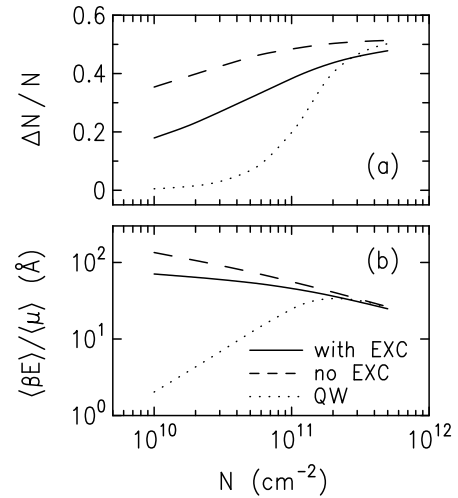}}
\vspace{3mm}
\caption[]{\label{fig:beta} (a) Spin splitting $\Delta N / N$ and
(b) effective spin splitting coefficient $\langle \beta^h_1 E_z
\rangle / \langle\mu_h\rangle$ as a function of $N$ for a 2D HH
system in the accumulation layer of a GaAs-Al$_{0.5}$Ga$_{0.5}$As
single heterostructure on a (001) GaAs substrate, calculated
including exchange-correlation (EXC, solid lines) and neglecting
exchange-correlation (dashed lines). For the dotted lines see text.}
\end{figure}

In Fig.\ \ref{fig:beta}(a) we show $\Delta N / N$ calculated as a
function of $N$ for a 2D HH system in the accumulation layer of a
GaAs-Al$_{0.5}$Ga$_{0.5}$As single heterostructure on a (001) GaAs
substrate. \cite{nminor} From $N=5\times 10^{11}$~cm$^{-2}$ to
$1\times 10^{10}$~cm$^{-2}$ the parameter $r_s$, the Coulomb energy
to Fermi energy ratio, increases from 4.3 to 17. Therefore, one can
expect that many-particle effects are quite important in this regime
of densities $N$. Indeed, we find that taking into account
exchange-correlation (solid lines) reduces $\Delta N / N$ as opposed
to a calculation without exchange-correlation (dashed lines). This
behavior, which is opposite to 2D electron systems, \cite{che99} can
be traced back to the fact that exchange-correlation increases the
subband spacings \cite{ste84} so that the Rashba coefficient
$\beta^h_1$ is reduced, in agreement with
Eq.~(\ref{eq:rash_HH_fak}).

It is convenient to characterize our numerical results in terms of
an effective Rashba coefficient $\langle \beta^h_1 E_z \rangle$
(Ref.\ \onlinecite{braket}). Assuming that the spin-split HH subband
dispersion is approximately of the form ${\cal E}_\pm^h (k_\|) =
\langle \mu_h \rangle k_\|^2 \pm \langle \beta^h_1 E_z \rangle
k_\|^3$, where $\mu_h$ (times $2/\hbar^2$) is the reciprocal
effective mass, we have
\widetext
\begin{equation}
\label{eq:beta}
\langle \beta^h_1 E_z \rangle =
\sqrt{\frac{2}{\pi}} \, \langle\mu_h\rangle \; \frac{
    N      \left(\sqrt{N + \Delta N} - \sqrt{N - \Delta N}\right)
+ \Delta N \left(\sqrt{N + \Delta N} + \sqrt{N - \Delta N}\right)}
{6\, N^2 + 2\, \Delta N^2}.
\end{equation}
\narrowtext\noindent
Figure \ref{fig:beta}(b) shows that $\langle \beta^h_1 E_z \rangle /
\langle\mu_h\rangle$ increases when $N$ is reduced.

\begin{figure}
\centerline{\includegraphics[width=0.67\columnwidth]{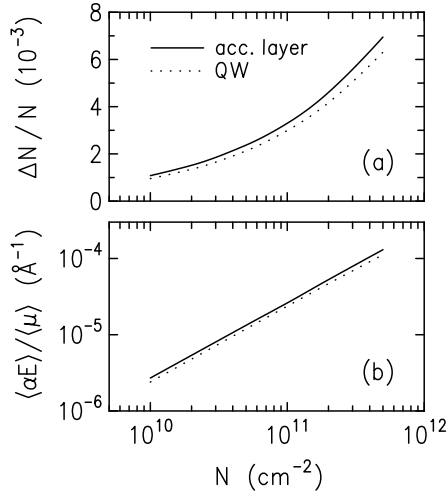}}
\vspace{3mm}
\caption[]{\label{fig:alpha} (a) Spin splitting $\Delta N / N$
and (b) effective spin splitting coefficient $\langle \alpha_1 E_z
\rangle / \langle\mu_c \rangle$ as a function of $N$ for a 2D
electron system in the accumulation layer of a
Ga$_{0.47}$In$_{0.53}$As-Al$_{0.47}$In$_{0.53}$As single
heterostructure (solid lines). For the dotted lines see text.}
\end{figure}

For comparison, we have calculated $\Delta N / N$ for a 2D electron
system in the accumulation layer of a
Ga$_{0.47}$In$_{0.53}$As-Al$_{0.47}$In$_{0.53}$As single
heterostructure \cite{nminor} [Fig.\ \ref{fig:alpha}(a)]. Here spin
splitting is given by Eq.\ (\ref{eq:rashbafull}). Therefore, the
spin-split subband dispersion is approximately of the form ${\cal
E}_\pm^c (k_\|) = \langle \mu_{\rm c} \rangle k_\|^2 \pm \langle
\alpha_1 E_z \rangle k_\|$, and we obtain similarly to Eq.\ 
(\ref{eq:beta}) (Ref.~\onlinecite{eng:approx})
\begin{equation}
\label{eq:alpha}
\langle \alpha_1 E_z \rangle = \sqrt{2\pi} \, \langle\mu_c\rangle
\big(\sqrt{N + \Delta N} - \sqrt{N - \Delta N}\big) .
\end{equation}
In Fig.\ \ref{fig:alpha}(b) it can be seen that, in contrast to
Fig.\ \ref{fig:beta}(b), the spin splitting coefficient $\langle
\alpha_1 E_z \rangle / \langle\mu_c\rangle$ decreases rapidly with
decreasing $N$. We remark that unlike the HH system in Fig.\ 
\ref{fig:beta}, exchange correlation has only a weak influence on
the electron system in Fig.\ \ref{fig:alpha} (Ref.\ 
\onlinecite{exc_imp}).

To further analyze our numerical results, we define an effective
electric field $\langle E_z\rangle = \langle \partial_z V_H (z)
\rangle$, where $V_H (z)$ is the Hartree potential without the
effective potential due to the position-dependent band edges. In an
accumulation layer the contribution of the space charge layer to
$V_H(z)$ is very small. \cite{ste74} It follows then, by partial
integration of the Poisson equation, that $\langle E_z\rangle = e /
(2\varepsilon\varepsilon_0) \, N$, where $\varepsilon$ is the
dielectric constant and the expectation value refers to the 2D
charge density that gives rise to $V_H (z)$. Using these values for
$\langle E_z\rangle$ and $\langle\mu_c\rangle = 89$~eV\AA$^2$ we
obtain $\langle\alpha_1\rangle \approx 34.3$~e{\AA}$^2$ independent
of $N$, consistent with Eq.\ (\ref{eq:rash_6c_sub_par}). This
implies that in Fig.\ \ref{fig:alpha}(b) the drastic change of
$\langle \alpha_1 E_z \rangle / \langle\mu_c\rangle$ merely reflects
the change of the electric field $\langle E_z \rangle$. On the other
hand, the weak variation of $\langle \beta^h_1 E_z \rangle /
\langle\mu_h\rangle$ in Fig.\ \ref{fig:beta}(b) indicates that the
``bare'' Rashba coefficient $\langle \beta^h_1\rangle$ increases by
a factor of 250 when $N$ is lowered from $5\times 10^{11}$ to
$1\times 10^{10}$~cm$^{-2}$ (Ref.\ \onlinecite{mhole}). This is in
good, qualitative agreement with the analytical model discussed
above that predicts an increase of $\beta^h_1$ by a factor of
$50^{4/3}\approx 184$. Note that for low densities the third order
perturbation approach, that underlies Eqs.\ 
(\ref{eq:rash_6c_sub_par}) and (\ref{eq:rash_HH_fak}), breaks down
because the subbands are merging together so that higher order
corrections become important. These higher order terms are fully
taken into account in our numerical calculations.
\cite{win00a,win93a} If the density $N$ is reduced below $10^{10}$
cm$^{-2}$ the Hartree potential and spin splitting are ultimately
controlled by the fixed concentration of minority impurities.
\cite{ste74,nminor}

It is interesting to compare the spin splittings in accumulation
layers with those in QW's where $E_z$ is tuned externally, e.g., by
means of gates. \cite{pap99} The dotted lines in Figs.\ 
\ref{fig:beta} and \ref{fig:alpha} show the calculated results for a
$200$~{\AA} wide rectangular QW where the external electric field
$E_z^{\rm ext}$ was chosen according to $ E_z^{\rm ext} (N) = e /
(2\varepsilon\varepsilon_0) \, N$. In an electron system (Fig.\ 
\ref{fig:alpha}) we obtain spin splittings very close to the results
for the accumulation layer. In particular, we have
$\langle\alpha_1\rangle \approx 30.6$~e{\AA}$^2$ independent of $N$.
Similarly, for a 2D HH system in a QW (Fig.\ \ref{fig:beta}) and $N
\lesssim 1\times 10^{11}$~cm$^{-2}$ we obtain
$\langle\beta^h_1\rangle \approx 7.54\times10^6$~e{\AA}$^4$. [For
larger densities higher order corrections in ${\cal E}_\pm^h (k_\|)$
become important. \cite{win00a}] Since in QW's the subband spacings
are essentially determined by the QW width (i.e., are independent of
$N$) this is consistent with Eq.\ (\ref{eq:rash_HH_fak}). These
calculations also indicate that for 2D HH systems in a QW, spin
splitting becomes negligible in the regime of low densities,
\cite{win00a} which is due to the fact that spin splitting of ${\cal
E}_\pm^h (k_\|)$ is proportional to $k_\|^3$. However, for 2D HH
systems in single heterostructures, spin splitting can be very
important in the low-density regime. We note that inversion layers
give results similar to QW's, but the specific numbers depend on the
details of the doping profile.

\begin{figure}
\centerline{\includegraphics[width=0.74\columnwidth]{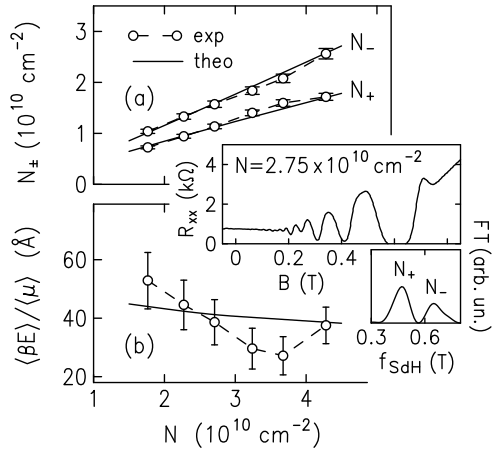}}
\vspace{3mm}
\caption[]{\label{fig:noh} Measured (circles) and calculated
(solid lines) spin subband densities $N_\pm$ (a) and effective spin
splitting coefficient $\langle \beta^h_1 E_z \rangle /
\langle\mu_h\rangle$ (b) as a function of density $N=N_+ + N_-$ for
a 2D HH system at a GaAs-Al$_{0.3}$Ga$_{0.7}$As single
heterostructure on a nominally undoped (311)A GaAs substrate with a
weak $p$-type background doping. The inset shows the measured
magnetoresistance $R_{xx}$ as a function of magnetic field $B$
(upper part) and the Fourier transform (FT) of $R_{xx}$ (lower part)
for $N=2.75\times 10^{10}$~cm$^{-2}$.}
\end{figure}

In order to reinforce our conclusions, we present next a comparison
between measured and calculated spin splittings in a
GaAs-Al$_{0.3}$Ga$_{0.7}$As single heterostructure grown on a
nominally undoped (311)A GaAs substrate with a weak $p$-type
background doping. \cite{nminor} A back gate was used to tune the
density $N$ from $1.8 \times 10^{10}$ to $4.2 \times
10^{10}$~cm$^{-2}$. To measure the spin subband densities $N_\pm$,
the Shubnikov-de Haas (SdH) oscillations at low magnetic fields $B$
were examined \cite{nit97,pap99} at a temperature $T\simeq 50$~mK
(see inset of Fig.\ \ref{fig:noh}). The frequencies $f_{\rm SdH}$ of
these oscillations are a measure of the zero-$B$ spin splitting.
\cite{ano_sdh} In Fig.\ \ref{fig:noh}(a) we present the measured and
calculated spin subband densities exhibiting remarkably close
agreement. Fig.\ \ref{fig:noh}(b) shows $\langle \beta^h_1 E_z
\rangle / \langle\mu_h\rangle$ determined by means of Eq.\ 
(\ref{eq:beta}). On average, $\langle \beta^h_1 E_z \rangle /
\langle\mu_h\rangle$ increases as the density is reduced. Taking
into account the orders-of-magnitude change that we have for
$\langle \beta^h_1 E_z \rangle / \langle\mu_h\rangle$ in QW's and
for $\langle \alpha_1 E_z \rangle / \langle\mu_c\rangle$ in electron
systems, the agreement between experiment and theory is quite
satisfactory. \cite{exp_err} We wish to emphasize that it is indeed
the anomalous enhancement of the Rashba coefficient in 2D HH systems
in accumulation layer-like single heterostructures that allows us to
experimentally resolve the spin splitting in this density regime;
data on QW samples with comparable densities reveal no measurable
spin splitting. \cite{tut01}

Support from NSF, DOE, and Humboldt Foundation are gratefully
acknowledged.

\vspace{-3mm}

\end{multicols}

\vspace{-6mm}
\begin{thebibliography}{10}
\vspace{-17mm}
  
 \bibitem{kit63} C. Kittel, {\em Quantum Theory of Solids} (Wiley,
  New York, 1963).
  
 \bibitem{ros89} U. R\"ossler, F. Malcher, and G. Lommer, in {\em
  High Magnetic Fields in Semiconductor Physics II}, edited by G.
  Landwehr (Springer, Berlin, 1989), p.\ 376.
  
 \bibitem{byc84} Y.~A. Bychkov and E.~I. Rashba, J.\ Phys.~C: Solid
  State Phys. {\bf 17}, 6039 (1984).
  
 \bibitem{nit97} J. Nitta {\it et~al.}, Phys.\ Rev.\ Lett. {\bf 78},
  1335 (1997).
  
 \bibitem{pap99} S.~J. Papadakis {\it et~al.}, Science {\bf 283},
  2056 (1999).
  
 \bibitem{pasps} Proceedings of the {\em International Conference on
  the Physics and Applications of Spin-Related Phenomena in
  Semiconductors} (Tohoku University, Sendai, 2000), to appear in
  Physica E.
  
 \bibitem{ste72} F. Stern, Phys.\ Rev.~B {\bf 5}, 4891 (1972).
  
 \bibitem{che99} G.-H. Chen and M.~E. Raikh, Phys.\ Rev.~B {\bf 60},
  4826 (1999).
  
 \bibitem{bir74} G.~L. Bir and G.~E. Pikus, {\em Symmetry and
  Strain-Induced Effects in Semiconductors} (Wiley, New York, 1974).
  
 \bibitem{tre79} H.-R. Trebin, U. R\"ossler, and R. Ranvaud, Phys.\ 
  Rev.~B {\bf 20}, 686 (1979).
  
 \bibitem{iv_rash} Eq.\ (\ref{eq:rash_6c_sub_par}) is essentially
  equivalent to Eq.\ (3.102a) in E. L. Ivchenko and G. E. Pikus,
  {\em Superlattices and Other Heterostructures} (Springer, Berlin,
  1997). The latter equation describes Rashba spin splitting in a
  quasi bulk material.
  
 \bibitem{las85} R. Lassnig, Phys.\ Rev.~B {\bf 31}, 8076 (1985).
  
 \bibitem{and97} E.~A. {de Andrada e Silva}, G.~C. {La Rocca}, and
  F. Bassani, Phys.\ Rev.~B {\bf 55}, 16293 (1997).
  
 \bibitem{win00a} R. Winkler, Phys.\ Rev.~B {\bf 62}, 4245 (2000).
  
 \bibitem{koster} We use the notation of G.~F. Koster {\em et al.},
  {\em Properties of the Thirty-Two Point Groups} (MIT, Cambridge
  MA, 1963).
  
 \bibitem{jor90} S. Jorda and U. R\"ossler, Superlatt. Microstruct.
  {\bf 8}, 481 (1990).
  
 \bibitem{pgroup} For the invariant expansion \cite{bir74} of the
  Hamiltonian it is irrelevant that the electric field further
  reduces the symmetry from $D_{2d}$ to $C_{2v}$.
 
 \bibitem{ref_ele} For the electron Hamiltonian
  (\ref{eq:rashbafull}) a refined analysis based on the point group
  $D_{2d}$ is not necessary because $T_d$ and $D_{2d}$ give the same
  results.
  
 \bibitem{lut56} J.~M. Luttinger, Phys.\ Rev. {\bf 102}, 1030
  (1956).
  
 \bibitem{ste74} F. Stern, Phys.\ Rev.\ Lett. {\bf 33}, 960 (1974).
  
 \bibitem{win93a} R. Winkler and U. R\"ossler, Phys.\ Rev.~B {\bf
  48}, 8918 (1993).
  
 \bibitem{ste84} F. Stern and S. {Das Sarma}, Phys.\ Rev.~B {\bf
  30}, 840 (1984).
  
 \bibitem{dft} A refined approach requires a spin density-functional
  theory for the eight-component spinors that combines many-particle
  effects and spin-orbit interaction. We assume, however, that the
  approach used here gives qualitatively correct trends.
  
 \bibitem{nminor} For a concentration of charged minority impurities
  $N_{\rm min} \lesssim 5 \times 10^{13}$~cm$^{-3}$ our results are
  essentially independent of $N_{\rm min}$.
  
 \bibitem{braket} We use angular brackets to indicate that the
  numerical calculation involves an averaging over
  position-dependent quantities.
  
 \bibitem{eng:approx} Equation (\ref{eq:alpha}) differs from the
  approximate results by G.\ Engels {\em et al.}\ [Phys.\ Rev. B
  {\bf 55}, R1958 (1997)]. The latter results are correct only up to
  first order of $\Delta N$.
  
 \bibitem{exc_imp} It was shown in Ref.\ \onlinecite{che99} that the
  exchange-induced enhancement of the Rashba coefficient $\alpha$
  becomes important for $r_s \gtrsim 8$. In 2D electron systems we
  have the largest Rashba spin splitting for semiconductors such as
  InAs with a small effective mass (i.e., small $r_s$).  For typical
  densities we have $r_s \lesssim 3$ so that usually many-particle
  corrections can be neglected for $\alpha$ in 2D electron systems.
  For the system in Fig.\ \ref{fig:alpha} we have $r_s = 0.5$ to
  $3.4$.
  
 \bibitem{mhole} According to our numerical calculations $\langle
  \mu_h\rangle$ decreases from 19.1 to 10.7~eV{\AA} for $N$ from
  $1\times 10^{10}$ to $5\times 10^{11}$~cm$^{-2}$.
  
 \bibitem{ano_sdh} This association may not be exact, i.e., $f_{\rm
  SdH}$ multiplied by $(e/h)$ can deviate slightly from the spin
  subband densities [R. Winkler {\em et al.}, Phys. Rev. Lett. {\bf
  84}, 713 (2000)].
  
 \bibitem{exp_err} We estimate that the experimental error in
  $N_\pm$ is of the order of $\pm 4$\%, giving an error in $\Delta
  N$ and $\langle \beta^h_1 E_z \rangle / \langle\mu_h\rangle$ of
  the order of $\pm 20$\%.
  
 \bibitem{tut01} E. Tutuc {\it et~al.}, Phys.\ Rev.\ Lett. {\bf 86},
  2858 (2001).

\end{thebibliography}
\end{document}